\documentclass[twocolumn,showpacs,preprintnumbers,amsmath,amssymb]{revtex4}
\usepackage{tabularx,graphicx}\begin{document}
\newcommand{\beq}{\begin{equation}}
\newcommand{\eeq}{\end{equation}}
\newcommand{\beqn}{\begin{eqnarray}}
\newcommand{\eeqn}{\end{eqnarray}}
\newcommand{\bmath}{\begin{subequations}}
\newcommand{\emath}{\end{subequations}}
\title{Hole superconductivity in Arsenic-Iron compounds}
\author{F. Marsiglio$^{a}$ and J. E. Hirsch$^{b}$ }
\address{$^{a}$Department of Physics, University of Alberta, Edmonton,
Alberta, Canada T6G 2J\\
$^{b}$Department of Physics, University of California, San Diego,
La Jolla, CA 92093-0319}
\begin{abstract}

Superconductivity above $25K$, and possibly above $40K$, has recently been discovered in $LaO_{1-x}F_xFeAs$ and related compounds.
We propose that this is another example
 of the mechanism of hole superconductivity at play. This requires the existence of hole carriers at the Fermi energy,
 which appears to contradict current observations. We propose that two-band conduction is taking place in these materials, that the negative ion $As^{-3}$ plays
 a key role,  and that superconductivity is non-phononic and driven by
 pairing and undressing of heavily dressed hole carriers to lower their kinetic energy. We
make several predictions of future observations based on our theory.

   \end{abstract}
\pacs{}
\maketitle

\section{Introduction}
Arsenic-iron compounds have just joined\cite{asfe} the ever-growing number of superconductors that do not appear to conform to the conventional BCS-Eliashberg electron-phonon explanation
of superconductivity\cite{parks}. Superconductivity above
$25K$ has been reported in  electron-doped $LaO_{1-x}F_xFeAs$\cite{asfe,zhu,chen,sefat,yang}, in hole-doped  $La_{1-x}Sr_xOFeAs$\cite{holedoped}, and very recently superconductivity above $40K$ in
electron-doped $SmFeAsO_{1-x}F_x$\cite{40k} and $CeO_{1-x}F_xFeAs$\cite{40k2}.
Eliashberg calculations\cite{boeri} based on density functional theory\cite{singh} as well as dynamical mean field theory calculations\cite{kotliar} appear to
rule out the electron-phonon mechanism as the cause of superconductivity in $LaO_{1-x}F_xFeAs$ and related compounds even at $25K$.
What then is the mechanism responsible for superconductivity? Where will the next high temperature superconductors be found?

The theory of hole superconductivity\cite{holemech} provides one   possible alternative to answer these questions (many other alternatives have been recently proposed\cite{singh,
kotliar, theories}),
 based on the same basic principles that we have  used to interpret the observed
superconductivity in cuprate oxides\cite{hole1}  (both hole-doped and electron-doped), and in $MgB_2$\cite{mgb2} : that high temperature superconductivity originates in
dressed hole carriers in negatively charged substructures that  pair and condense to lower their kinetic energy\cite{kinetic}. In the cuprates, the holes reside in $O^{2-}$ ions in the highly
negatively charged $Cu-O$ planes. In $MgB_2$, the holes reside in $B^-$ ions in the highly negatively charged $B$ planes. In the arsenic-iron compounds, we propose
that superconductivity is driven by holes in $As^{3-}$ ions that   conduct  in  highly negatively charged $Fe-As$ planes.

 Based on this theoretical perspective, we predict that positive Hall coefficient  will be measured in single crystals of $LaO_{1-x}F_xFeAs$
 $SmFeAsO_{1-x}F_x$ and $CeO_{1-x}F_xFeAs$ for in-plane transport;  observation
 of tunneling asymmetry of
universal sign\cite{tunn};  two-gap superconductivity, with the large gap associated with hole carriers and the small gap with electron carriers\cite{twoband};
optical spectral weight transfer from high to low frequencies as the temperature is lowered\cite{optical}; positive in-plane pressure dependence of $T_c$\cite{hole1}.
We also discuss how to achieve higher transition temperatures based on this theory.

Our point of view differs significantly from the prevalent point of view. The prevalent point of view maintains that superconductivity in $MgB_2$ is fully explained by conventional electron-phonon
theory; that superconductivity in the cuprates is driven by electrons in copper d orbitals, and that it has d-wave symmetry at least in the hole-doped case; that in the electron-doped cuprates the
dominant carriers are electrons rather than holes; that magnetic fluctuations arising from a strong Hubbard-U repulsion on copper ions
drives superconductivity and provide the 'glue' for Cooper pairing
in the cuprates. For arsenic-iron compounds, a consensus is rapidly growing that (in most compounds) the carriers are electron-like\cite{asfe,chen,zhu,sefat}, that they reside in  iron d-bands at the Fermi
energy\cite{boeri,singh,kotliar},
and that incipient ferromagnetism or antiferromagnetism of electrons in Fe orbitals may drive pairing
and superconductivity\cite{boeri,singh,kotliar,theories}. Based on that consensus one would $not$ predict the observations listed in the previous paragraph.

We argue that our point of view on the explanation of  high temperature superconductivity
 is compelling because it relies on a minimal number of unifying assumptions and because it is easily falsifiable, in contrast to the
prevalent point of view. (1) Conduction has to occur in substructures (planes) that have excess negative charge. If instead the planes have excess positive charge, or are charge neutral,
high temperature superconductivity will not occur. Neither electron-phonon theory nor magnetic-based theories can make such a general prediction.
(2) Hole carriers have to exist. Again,  neither electron-phonon theory nor magnetic-based theories can make such a general prediction. We made these predictions based on experimental evidence on hole-doped cuprates in 1989. Subsequent observations in
electron-doped cuprates and in $MgB_2$ have been fully consistent with these predictions. Arsenic-iron compounds are the next test of our theory:
given the observations to date these compounds  could potentially refute it. If they don't, they will provide additional evidence in its favor.

\section{nature of the charge carriers}
The Hall coefficient has been measured in polycrystalline samples of $LaO_{1-x}F_xFeAs$ by Zhu et al\cite{zhu}, Chen et al\cite{chen},
Sefat et al\cite{sefat}, and Yang et al\cite{yang}. In all cases it is found to be
negative and weakly temperature dependent, becoming more negative as the temperature decreases. However,
in a multi-band situation, the Hall coefficient will generally exhibit temperature dependence due to the different temperature dependent of the mobilities in the different bands.
So measurement of a negative Hall coefficient certainly does not imply absence of hole carriers.
For an isotropic two band model with electron and hole carriers of densities $n_e$ and $n_h$  the Hall coefficient is
\beq
R_H=-\frac{1}{n_e ec}\frac{1-(n_h/n_e)(\mu_h/\mu_e)^2}{1+(n_h/n_e)(\mu_h/\mu_e)^2}
\eeq
and will be negative if the mobility of the dressed hole carriers ($\mu_h$)  is much smaller than that of the electron carriers ($\mu_e$) , as we expect. However, we expect that
in-plane transport will start being dominated by  hole carriers as the temperature is lowered and the hole mobility increases. Just as in the case of electron-doped cuprates\cite{electrondoped},
we expect that measurement of the in-plane Hall coefficient in single crystal samples as the temperature is lowered will show  a trend towards zero and
a change in sign to positive before the onset of superconductivity.

Simple valence counting indicates that the $La^{3+}O^{2-}$     unit donates one electron to the $Fe^{2+}As^{3-}$ unit, so the substructure where the conduction is expected to
occur has one extra electron per $FeAs$ unit.  This is similar to the cuprates, where there are two extra electrons per $CuO_2$ unit in the plane donated by the
off-plane atoms, and to $MgB_2$ which has one extra electron per $B$ atom in the $B^-$ plane donated by the off-plane $Mg^{2+}$ ion. Thus, the first requirement for
high temperature superconductivity in our theory is clearly satisfied. The reason negatively charged substructures are helpful  is because they give rise to large
orbital relaxation effects when a hole travels through such a structure, which is the mechanism that leads to pairing of hole carriers in our model.

In the absence of doping, the compound $LaO Fe As$ exhibits superconductivity with $T_c$ around $3K$\cite{asfe}. When $O$ is substituted by $F$, electrons are
added to the $FeAs$ plane, and $T_c$ increases above $25K$. It has very recently been reported\cite{holedoped} that upon substituting $La$ by $Sr$, which would correspond
to hole doping, superconductivity with $T_c\sim 25K$ also occurs, namely  in $(La_{1-x}Sr_x)OFeAs$ with $x\sim 0.13$. How can this apparent electron-hole `symmetry' be understood
within the theory of hole superconductivity?

Fig. 1 shows the generic $T_c$ versus hole concentration dependence in the model of hole superconductivity. We assume that this band originates in direct overlap
of $As$ p-orbitals. The undoped material
is proposed to correspond to the underdoped regime where this band is almost full and
$T_c$ is almost zero. Hence the hole-doped case ($Sr$ substituting for $La$) is easy to understand: for each $Sr$ substituting $La$, one
hole gets added to the $As$ band and $T_c$ increases up to the maximum in Fig. 1. The
electron-doped case ($F$ substituting for $O$) is less straightforward: we need to assume that as electrons are added  both
electrons and holes are doped to the Fe-As planes. This would parallel the scenario that we proposed for the electron-doped cuprates\cite{edoped}.

\begin{figure}
\resizebox{8.5cm}{!}{\includegraphics[width=7cm, angle=-90]{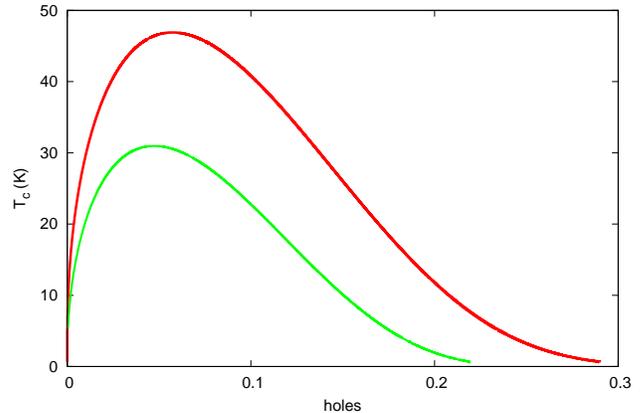}}
\caption {(Color online) $T_c$ versus hole concentration in the arsenic band for the two-band model
(solid red line) and the single-band model (dashed green line). For simplicity we used equal
bandwidths (1 eV) for both the arsenic-based and iron-based bands, with the center of the Fe
band shifted with respect to the center of the As band. We used interaction parameters $U_a = 5$
eV and $K_a = 1.86$ eV, and $U_{f} = K_{f} = 0$. The interband coupling was chosen to
be a constant, $V_{ad} = 0.2$ eV. For the single-band result we set $V_{ad} = 0$.}
\label{figure1}
\end{figure}

Schematically, as shown in Fig. 2, when an electron is added to the $Fe-As$ plane, we assume it goes onto the $Fe^{++}$ ion. It will repel neighboring electrons in $As^{3-}$ ions and push one electron into
another neighboring $Fe^{++}$. The net result would be that for each electron added to the $Fe-As$ plane, two $Fe^{++}$ ions get converted into $Fe^{+}$ and one $As^{3-}$ ion turns into
$As^{2-}$. Thus electron doping creates added electron carriers in a $Fe$ band, and added hole carriers in a purely $As$ band, which would drive the system superconducting just like in the
case of electron-doped cuprates.

  \begin{figure}
\resizebox{8.5cm}{!}{\includegraphics[width=7cm]{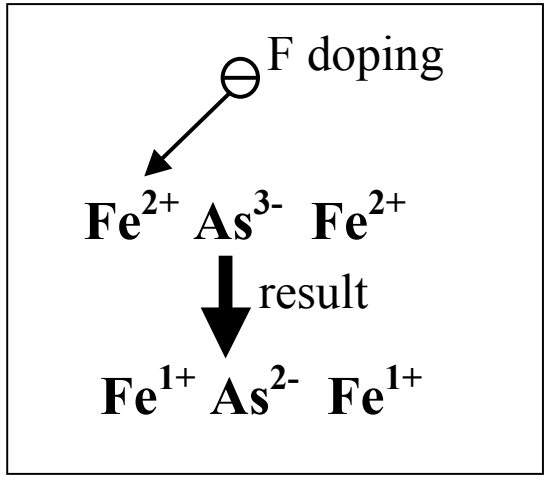}}
\caption {Schematic depiction of how holes are created by electron doping. The electron added to $Fe^{2+}$ repels an electron from $As^{3-}$ to the neighboring
$Fe^{2+}$, leaving behind a hole in arsenic ($As^{2-}$). }
\label{figure2}
\end{figure}

Thus, we propose that a `minimal' model to describe superconductivity in $As-Fe$ compounds has to contain two bands: a hole band derived from $As$ $p$ orbitals (for example $p_z$ orbitals
forming $\pi-$bonds) and a band involving $Fe$ $d$ orbitals, as shown schematically in Fig. 3. We expect the $As$ band to be very narrow because of the band narrowing effect due to
orbital relaxation. Furthermore, its bandwidth will increase with hole doping (this effect is not shown in Fig. 3)\cite{holemech}. Upon hole doping the Fermi level will move down, adding holes to (removing electrons from)
both the $As$ and $Fe$ bands (Fig. (3b)). Upon electron doping however there should also be a relative shift of the position of the  two bands, with the
$As$ band moving up  in energy and the $Fe$  band moving down in energy, as shown in Fig. (3c), resulting in $hole$ doping of the $As$ band and additional electron doping of the $Fe$ band.

This band shift results
self-consistently from the
self-doping process discussed in connection with Fig. (2): as  electrons are pushed out from  $As^{3-}$ ions onto $Fe^{2+}$ ions,
the (negative) carriers in the $Fe$ band feel a higher electric potential
from the more positively charged $As$ ions and hence their energy is lowered. Conversely, the carriers in the $As$ band feel a lower electric potential because the $Fe$ ions have become more negatively charged,
hence their energy is raised. This results in the relative band shifts shown in Fig. (3c). For this process to take place requires the `charge transfer gap'\cite{ctgap} (the energy cost in transferring
an electron from $As^{3-}$ to $Fe^{2+}$) to be small, as in the case of the $T'$ structures in the cuprates that allow for electron doping in the $Cu-O$ planes\cite{edoped}.

Thus we propose that a rigid band model is approximately valid for the case of hole doping of $As-Fe$ planes in these structures
(except for the neglected band expansion effect) but is qualitatively wrong for  electron doping.
\begin{figure}
\resizebox{8.5cm}{!}{\includegraphics[width=7cm]{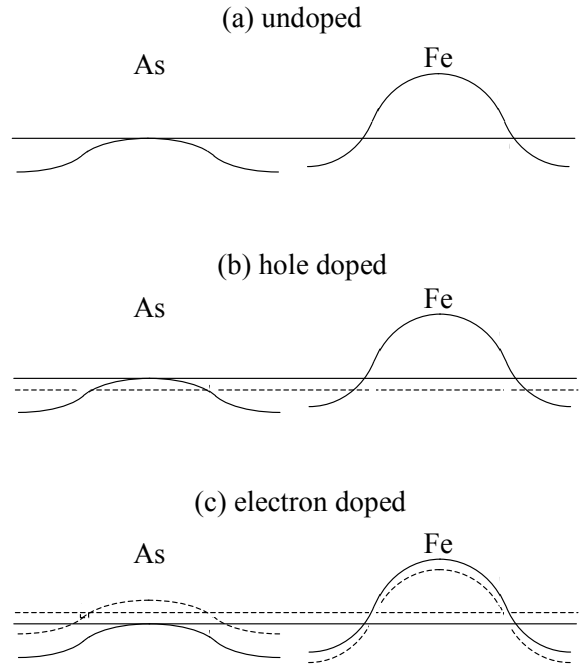}}
\caption {Schematic depiction of $As$ and $Fe$ bands. In the undoped case (a), the Fermi level (horizontal full line) is very close to the top of the $As$ band since $T_c$ is very low.
Upon hole doping ((b)), the Fermi level moves down (dashed horizontal line in (b)) and the bands don't move. Upon electron doping ((c)), the Fermi level moves up (dashed horizontal line),
the $As$ band moves up and the $Fe$ band moves down, as indicated by the dashed bands in (c). }
\label{figure2}
\end{figure}

\section{two-band superconductivity}

Our model for the arsenic-iron compounds utilizes two bands, one of primarily arsenic
character, and the other of primarily iron character, as depicted in Fig. (3). The reduced Hamiltonian to describe such a model, following Suhl et al. \cite{suhl59}, is
\begin{eqnarray}
& & H = \sum_{k\sigma} (\epsilon_{k}^a - \mu)a_{k\sigma}^\dagger a_{k\sigma} +
\sum_{k\sigma} (\epsilon_{k}^d - \epsilon_0 - \mu)d_{k\sigma}^\dagger d_{k\sigma} +
\nonumber \\
& &\sum_{k k^\prime} V_{kk^\prime}^{aa} a_{k\uparrow}^\dagger a_{-k\downarrow}^\dagger
a_{-k^\prime\downarrow} a_{k^\prime\uparrow}
+
\sum_{k k^\prime} V_{kk^\prime}^{dd} d_{k\uparrow}^\dagger d_{-k\downarrow}^\dagger
d_{-k^\prime\downarrow} d_{k^\prime\uparrow}
+
\nonumber \\
& & \sum_{k k^\prime} V_{kk^\prime}^{ad} \bigl( a_{k\uparrow}^\dagger a_{-k\downarrow}^\dagger
d_{-k^\prime\downarrow} d_{k^\prime\uparrow} + d_{k\uparrow}^\dagger d_{-k\downarrow}^\dagger
a_{-k^\prime\downarrow} a_{k^\prime\uparrow}\bigr).
\label{ham}
\end{eqnarray}
As discussed in Ref. \cite{twoband} we retain the simplest interband interaction, and in
what follows adopt a constant interband potential: $V_{kk^\prime}^{ad} = V_{ad}$. We have
used a hole notation, so that the $a^\dagger$ and $d^\dagger$ operators correspond to hole creation
operators in the As and Fe band, respectively, and similarly for the annihilation operators.
We adopt a flat density of states for both bands, each with bandwidth $D_i$. The single particle
energies are measured from the center of each band, and the Fe band is shifted by an amount
$\epsilon_0$ with respect to the As band.

The intraband potentials are assumed to have identical form; we adopt the form from Ref. \cite{holemech}:
\begin{equation}
V_{kk^\prime}^{ii}=U_i + K_i \biggl({\epsilon_k^i \over D_i/2} + {\epsilon_{k^\prime}^i \over D_i/2}
\biggr) + W_i {\epsilon_k^i \over D_i/2} {\epsilon_{k^\prime}^i \over D_i/2},
\label{pot}
\end{equation}
where $U_i$ corresponds to the on-site repulsion, $K_i$ the modulated hopping,
and $W_i$ the nearest-neighbor repulsion. These interactions lead to a BCS ground state that is
superconducting, and an (s-wave) order parameter with the form
\begin{equation}
\Delta_i(\epsilon) = \Delta_i^m\bigl(c_i - {\epsilon \over D_i/2}\bigr),
\label{gap}
\end{equation}
as found previously \cite{twoband,holemech}. Further details are available in these references.

The generic behavior of $T_c$ as a function of doping was already displayed in Fig. (1). The
existence of interactions in two bands (instead of a single band) does not alter the behavior of $T_c$ in any significant manner. For simplicity we chose the nearest neighbor repulsion equal
to zero for calculations in this paper; in Fig. (1) $U_a = 5$ eV, $K_a = 1.86$ eV, and
$U_d = K_d = 0$. The magnitude of $T_c$ is of course determined by the relative strength of the interactions; it is noteworthy that with our choice the interaction is effectively repulsive
everywhere in the band (see Fig. 8 of Ref. \cite{holemech}).

In Fig. (4) we show the result of self-consistent calculations of the gap, as a function of
temperature. We chose parameters as in Fig. (1), with a hole concentration corresponding to the
maximum $T_c$. Note that the gap is determined by minimizing the energy dispersion relation.
Provided we are not too close to the band edges (as is the case here) the result is
\begin{equation}
\Delta_{0i} = { \Delta_i(\mu + \epsilon_0^i) \over \biggl( 1 + \bigl({\Delta_i^m \over
D_i/2}\bigr)^2 \biggr)^{1/2}}.
\label{gaps}
\end{equation}
In Fig. (4) these are plotted for the As band (solid, red curve) and the Fe band (dashed, green
curve). They both display the generic BCS gap temperature dependence; clearly the As band drives
the superconductivity, and the Fe superconducts only because of interband interactions. A generic
feature is that the zero temperature gap ratio is larger (smaller) than the weak coupling BCS
result for the As (Fe) gap.

\begin{figure}
\resizebox{8.5cm}{!}{\includegraphics[width=7cm, angle=-90]{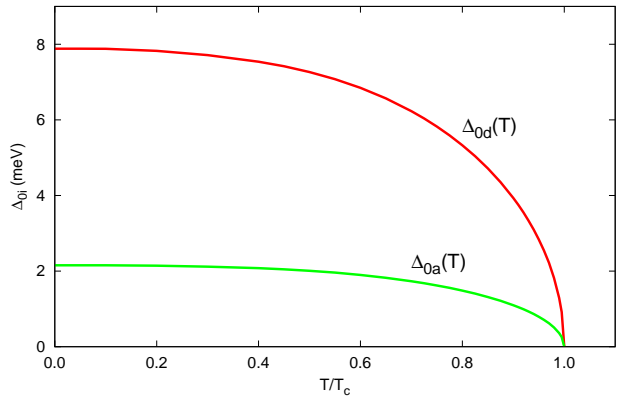}}
\caption {(Color online) The As-based gap, $\Delta_{0a}$, and the Fe-based gap, $\Delta_{0d}$ versus
temperature. The temperature dependence is essentially identical, and very
BCS-like. For these parameters, the gap ratios at zero temperature are $3.9$ and $1.1$ for
the large and small gap, respectively. The ratio between the two is $\Delta_{0a}/\Delta_{0d} = 3.7$.}
\label{figure4}
\end{figure}

The gaps can most readily be measured in tunneling experiments. In Fig. (5) we plot the expected
$dI/dV$ vs. sample voltage, for the same parameters as in Fig. (4). In (a) we used underlying
electronic densities of states which were identical for both the As and the Fe bands. The presence
of two gaps is readily apparent in the figure, especially at low temperatures. The relative sizes
of the gaps are unknown at present. Here, they were determined by our choice of no intraband interactions in the Fe band, but of course some (presumably repulsive) interactions exist, and this
would decrease the smaller gap in comparison to the large one.

In Fig. (5b) we show results for the same parameters as in (a), except that we have included an
extra degeneracy in the Fe band (by a factor of 3). This is expected from naive considerations
of the orbital degeneracy in Fe vs. As. Note that this effect alters the relative weight of the coherence peaks associated with the smaller gap compared to the peaks associated with the larger
gap. In either case the two gap structure should be evident in the tunneling. Finally, note that
the larger gap coherence peaks display an asymmetry; injecting holes into the sample (negative bias)
results in a   higher coherence peak than removing holes (positive bias) \cite{tunn} (note that we reversed our sign convention from our older references to conform with experimental
practice). This asymmetry of universal sign  is a direct consequence of the form of the order parameter, Eq. (\ref{gap}).

\begin{figure}
\resizebox{8.5cm}{!}{\includegraphics[width=7cm]{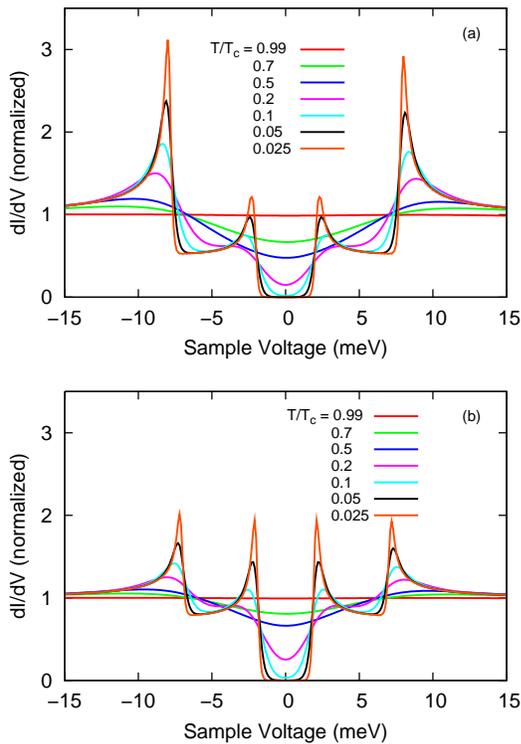}}
\caption {(Color online) $dI/dV$ versus sample voltage, for the parameters used in Fig. (4),
for a variety of temperatures below $T_c$. In (a) we use identical electronic densities of states
in the As and Fe bands. In (b) we increase the degeneracy in the Fe band by a factor of 3.
In both cases the two gap structure in the tunneling curves is most evident at low temperatures,
and a tunneling asymmetry exists --- large gap coherence peak higher on the left than on the right.}
\label{figure5}
\end{figure}

\section{the key role of $As^{3-}$}

In the theory of hole superconductivity the atomic charge  parameter  $Z$, where $(Z-2)$ is the charge of the negatively charged ion when the relevant band is full, plays a key role\cite{molecules}.
As $Z$ decreases the degree of orbital relaxation when a hole is added to the negative ion increases, causing an increase in the pairing interaction strength
arising from hopping renormalization as well as a decrease in the effective on-site repulsion. Both of these effects contribute to a higher $T_c$.
For the classes of materials discussed here $MgB_2$ corresponds to $Z=1$ (charge of $B^+$), the cuprates to $Z=0$ ($O^o$) and the $Fe-As$ materials to
$Z=-1$ ($As^{1-}$). Thus, the latter class of materials has the potential of yielding the highest  $T_c$'s.
This was explicitly predicted in ref.\cite{nitrogen} where we remarked:
``Even stronger hole dressing and higher $T_c$Õs would be expected
in a structure with even smaller Z $-$ for example, if
one managed to make a material with $N^{3-}$  planes doped with
some holes ($Z=-1$).''

Band structure calculations using density functional theory (DFT) place the bands arising from $As$ p-orbitals around $3eV$ below the Fermi energy\cite{singh},
while the bands crossing the Fermi energy arise from  $Fe$-d orbitals.  However,
we require that hole doping results in hole carriers in a nearly full band that arises from direct overlap of $As$ p-orbitals. Do the band structure calculations
rule out our picture?

We propose that the band structure calculations are in error when it comes to predicting the location of bands arising from overlap of orbitals of highly negatively charged
ions. This is because DFT does not accurately take into account the effect of local orbital relaxation, which becomes increasingly important the more
negative the ion is. Consider the process of removing an electron from (i.e. adding a hole to) an $As^{3-}$ ion. The electrons in the resulting
$As^{2-}$ ion will have their orbits contracted and relaxed to a lower energy state. To the extent that DFT does not consider (or underestimates) this local
relaxation process it will overestimate the energy of the final state, and hence overestimate the energy cost of adding a hole to $As^{3-}$.
For example, for the simple case of the $H^-$ ion the energy cost of creating a hole (ionization energy) assuming the electron in the final state does not relax
is $1.3 eV$ higher than when orbital relaxation is allowed in a Hartree calculation\cite{nitrogen}. Thus the calculated band from  such an orbital ignoring the
relaxation effect would be placed $1.3 eV$ $lower$ than
where it really is.
Because $As^{3-}$ is highly negatively charged the electronic states are changed significantly in the different ionization states, and the
energy difference between the relaxed and unrelaxed states should be large. Thus it is plausible that a DFT calculation that does not take
this effect into account will place the $As$ bands several $eV$ lower than where they really are, and that in fact adding holes to the undoped
compound may  put holes into a pure $As$ band.

What  about the electron-doped case? First, it requires a structure where the Madelung energy of $As^{3-}$ and $Fe^{2+}$ is similar, as in the case of
electron-doped cuprates\cite{edoped}. Furthermore, the same orbital relaxation effect discussed above will make it easier for an electron doped onto a $Fe^{2+}$ ion to
repel an electron out of a  neighboring $As^{3-}$ ion and create a hole in it.  Thus, just like in the case of electron-doped cuprates\cite{edoped},
adding electrons to the conducting planes would result in twice as many electrons added to an electron band and holes induced in the hole band arising from the
negative ions. By not properly taking into account the local orbital relaxation of the $As^{-3}$ ions,  DFT would miss this effect and predict that only electron carriers are created
upon electron doping.

Finally, we have argued that high $T_c$ superconductivity is favored when hole propagation occurs in substructures (planes) that are highly negatively charged\cite{mgb2}.
Assume the valence states in the $Fe-As$ planes corresponded instead to $Fe^{3+}-As^{-3-}$ ions, i.e. a charge neutral structure. In that case the
$Fe-As$ bond would be more covalent and the electrons would be located less on the negative ions and more on the covalent bond,
 the effect of orbital relaxation in the negative ion would be reduced, and we would not expect high $T_c$ superconductivity.
 Using this argument we predicted that superconductivity would $not$ occur in
$LiBC$, as had been predicted from electron-phonon theory\cite{pickett}, and indeed none has been found\cite{nopickett}.

\section{predictions}
Our point of view leads to the following predictions for future experimental observations:

\noindent $\bullet$Hall coefficient:
In the hole-doped samples, positive Hall coefficient has been measured, in accordance with expectations\cite{holedoped}.
In electron-doped samples, all measurements of Hall coefficient reported so far have been for polycrystalline samples and yielded a negative Hall
coefficient\cite{chen,zhu,sefat,yang}. However, for transport in the $a-b$ plane in single crystal samples
we predict that a positive Hall coefficient will be measured at low temperatures at least in the doping range where $T_c$ is highest. Furthermore even in regimes where the Hall coefficient
is negative there should be evidence of two-band conduction and that hole carriers start to dominate  the transport as the temperature is lowered, as is found
in electron-doped cuprates\cite{electrondoped}.

\noindent  $\bullet$ Tunneling: our theory predicts tunneling asymmetry of universal sign\cite{tunn} (larger peak for negatively biased sample) arising from the
hole band. In fact, point contact tunneling experiments already have shown this asymmetry\cite{tunnexp}. For tunneling in the ab plane we expect that evidence for two-band superconductivity,
as seen in $MgB_2$\cite{mgb2tunn}, will be seen, at least in the electron-doped materials. For the smaller gap we do not expect asymmetry in the tunneling peaks.

\noindent  $\bullet$ Isotope effect: we predict a positive isotope effect in $T_c$ for $As$ substitution but no isotope effect for $Fe$ or $O$ substitution\cite{mgb2}. A more difficult
experiment would be to measure the tunneling characteristic for As substitution; then a
direct identification of the large gap coherence peak with As could be made.
Our model also predicts a positive isotope effect in the superfluid weight (inverse penetration depth squared) due to the effective mass
reduction (increase in hopping amplitude) upon pairing. (This effect was first predicted by the bipolaronic theory of superconductivity)\cite{alexandrov}). However, in a multi-band situation if the superfluid
weight is dominated by carriers in the electron band this effect may be difficult to detect.

\noindent  $\bullet$ Pressure effect: our theory predicts increased $T_c$ as the distance between $As$ ions in the plane decreases, hence a positive pressure coefficient
for in-plane pressure. For pressure perpendicular to the planes the $T_c$ dependence on pressure should be weak and could be of either sign\cite{holemech}.

\noindent  $\bullet$ Optical sum rule violation: we predict optical spectral weight transfer from high frequencies to the zero frequency $\delta$-function as the system is cooled into the
superconducting state, with the effect being largest in the underdoped regime\cite{optical}.

\section{how to get higher $T_c's$}
The theory of hole superconductivity yields simple prescriptions to search for higher $T_c$ materials, which we summarize here.

\noindent $\bullet$ Holes conducting through an almost filled band arising from direct overlap of orbitals of negative ions are essential. The more negatively charged the ion is, the better,
so triply negatively charged anions in the fifth column (N, P, As, Sb, Bi) are best.

\noindent $\bullet$ Neither magnetic fluctuations nor d-electrons are needed, as illustrated by the case of $MgB_2$. Hence
the same structure with e.g. $Cd^{2+}$ ions instead of  $Fe^{2+}$ ions  could potentially yield high $T_c$.

\noindent $\bullet$ Negatively charged planes or other substructures where hole conduction occurs. The larger the negative charge the better.
Hence if $Fe^{2+}$ could be replaced by a $monovalent$ positive ion $T_c$ should be greatly increased.

\noindent $\bullet$ As small a distance between negative ions as  possible. This could be achieved by various off-plane substitutions with atoms of smaller size.

\section{discussion}

Our interpretation of the origin of superconductivity in arsenic-iron compounds differs substantially from other proposed explanations, and we certainly cannot argue at this point that
experiments provide compelling support for our point of view over others. However,
one strength of our approach is that it makes very definite predictions, and past experience shows that these predictions
are often confirmed much later. Examples are our  prediction of the existence of hole carriers in electron-doped materials (1989)\cite{edoped}, of tunneling asymmetry
of universal sign (1989)\cite{tunn}, of apparent optical sum rule violation and color change
(1992)\cite{apparent}, and of hole carriers in $MgB_2$ (early March 2001)\cite{mgb2}. Except for the latter that was confirmed only a few days later, the first three predictions were confirmed many years later, and in the
intervening years experiments appeared to be at odds with these predictions.

On the other hand, some predictions of our theory have not been experimentally confirmed, for example the prediction of positive pressure effect on $T_c$ in $MgB_2$\cite{mgb2}.
This can be understood by hypothesizing that other effects in addition to
reduction of the distance between atoms, such as charge transfer between different parts of the system, dominate when pressure is applied.

The greatest strength of our approach however is that it provides a unified point of view to understand superconductivity in all materials. As the number of distinct classes of superconducting
materials increases, a new parameter regime opens up as a testing ground for theories. It
becomes increasingly unlikely that a different mechanism is required for each new class. As in other
fields (e.g. particle physics) one instead searches for unifying ideas that explain a particular phenomenon in a variety of classes. As we have argued here, the Iron-Arsenic compounds provide yet
another class in which our proposal can be put to the test.

\begin{acknowledgments}

FM is grateful for support from
the Natural Sciences and Engineering Research Council of Canada (NSERC),
by ICORE (Alberta), and by the Canadian Institute for Advanced Research (CIfAR).

\end{acknowledgments}

 \end{document}